\documentclass[graphicx, nofootinbib, notitlepage, twocolumn, amsmath, amssymb, longbibliography,floatfix]{revtex4-2}
\usepackage{dcolumn}  
\usepackage{bm}       
\usepackage{physics}  
\usepackage{braket}
\usepackage{dsfont}
\usepackage{enumitem}
\usepackage{tikz}
\usepackage{xcolor}
\definecolor{BLUE}{rgb}{0,0,1}
\usepackage{siunitx}

\def\beit{\begin{itemize}}
\def\eit{\end{itemize}}

\DeclareSIUnit{\angstrom}{\textup{\AA}}
\newcommand{\rev}[1]{#1}
\newenvironment{revblock}{}{}

\begin{document}

\title{Hydrogen Chemisorption and Current-Induced Spin Polarization on NbP}
\author{Luis Martinez-Gomez}
\author{Raphael F. Ribeiro}
\email[]{raphael.ribeiro@emory.edu}
\affiliation{Department of Chemistry and Cherry Emerson Center for 
Scientific Computation, Emory University, Atlanta, GA, 30322}

\date{\today}


\begin{abstract}
\begin{revblock}
Topological semimetals have been proposed as electrocatalytic platforms because their surface states can connect adsorbate bonding with interfacial charge and spin responses. Here we investigate hydrogen chemisorption on NbP(001) using density functional theory and Wannier-based analyses of surface spectra, chemical bonding, and current-induced spin polarization. Comparing calculations with and without spin--orbit coupling allows us to examine nodal-line-derived surface states and Weyl Fermi arcs on the same surface. Spin--orbit coupling leaves the adsorption thermodynamics essentially unchanged but reorganizes how hydrogen bonds with the Nb- and P-derived surface orbitals. The characteristic Fermi-arc branches persist after adsorption, with H-derived spectral weight appearing within the same near-Fermi-level surface manifold. Together, these results show that chemisorbed hydrogen participates in the spin-textured surface electronic response. The accompanying adsorbate-local current-induced spin polarization points toward opportunities to connect surface chemistry with electrically controlled spin phenomena in topological semimetals.
\end{revblock}
\end{abstract}

\maketitle

\noindent



\section{Introduction}
Weyl semimetals (WSMs) are three-dimensional topological semimetals with pairs of bulk band crossings, or Weyl points, between conduction and valence bands. On a surface, projections of Weyl points with opposite chirality are connected by open constant-energy contours known as Fermi arcs \cite{yan2017topological, murakami2007phase, wan2011topological, weyl1929electron}. Since their prediction and experimental realization \cite{weng2015weyl,nielsen1983adler,wan2011topological, ruan2016symmetry, jia2016weyl,xu2015discovery,lv2015experimental, yang2015weyl, lv2015observation}, WSMs have attracted interest for optoelectronic, transport, and spintronic applications \cite{asadchy2020sub,garcia2020optoelectronic,wang2018electron,kharzeev2013anomaly,lundgren2014thermoelectric,hosur2012charge,Sun2016}. This interest is closely connected to their surface electronic structure: Fermi arcs are surface-localized boundary modes tied to bulk topology and, in materials with spin--orbit coupling (SOC), they carry nontrivial spin texture \cite{sun2015topological,xu2016spin}.

The TaAs-family compounds provide a controlled setting for comparing related surface-state manifolds on the same chemical surface \cite{lv2015experimental, sun2015topological, lee2015fermi, huang2015weyl}. Without SOC, these noncentrosymmetric semimetals host nodal rings and corresponding nodal-line-derived surface states;\rev{\cite{fang2016topological}} including SOC gaps the mirror-plane nodal rings and yields Weyl points connected by Fermi arcs.\cite{weng2015weyl,sun2015topological, yu2017nodal} The SOC on--off comparison therefore changes the topology, spin texture, and momentum-space structure of the surface states while preserving the local atomic framework.

Surface chemistry depends on how adsorbate orbitals hybridize with substrate states to form bonding and antibonding hybrid electronic states at the interface \cite{hammer2000theoretical}. Conventional catalytic descriptors summarize aggregate electronic contributions within ordinary band theory.\cite{medford2015sabatier, jiao2022descriptors} Resolving the relative roles of topological, trivial-surface, and bulk states in bonding on topological semimetals therefore provides an important momentum- and state-resolved extension.

This distinction is important for WSM surfaces because SOC Fermi arcs combine surface localization with momentum-dependent spin texture. If an adsorbate hybridizes with these states, the interface can acquire both H-derived spectral weight at selected momenta and an adsorbate-local spin response. Under an applied electric field, spin--momentum locking can generate an induced spin polarization, the nonequilibrium interfacial spin response commonly associated with the Edelstein effect \cite{johansson2024theory, aronov1989nuclear, dyakonov1971current, edelstein1990spin} and thereby influence local chemistry. This motivates a combined analysis of adsorption thermodynamics, momentum-resolved bonding, and adsorbate-projected electric field-induced spin polarization \cite{johansson2018edelstein, johansson2016theoretical}.

The hydrogen evolution reaction\cite{eftekhari2017electrocatalysts, zheng2015advancing, zhao2018heterostructures} (HER) provides a technologically relevant setting for this analysis. Its elementary adsorption and desorption steps are commonly summarized by the hydrogen adsorption free energy, $\Delta G_{H^\ast}$ \cite{parsons1958rate,trasatti1972work}. Platinum-group metals are highly active HER catalysts, although cost and scarcity motivate the search for alternatives \cite{zeradjanin2016critical, hammer2000theoretical, wu2020electronic, sarkar2018overview, hansen2021there}. Weyl semimetals have been proposed as one such class, with NbP showing strong activity within the TaAs family and computed hydrogen adsorption energetics consistent with favorable HER behavior \cite{rajamathi2017weyl}. The combination of high carrier mobility \cite{shekhar2015extremely}, a finite density of states near the Fermi energy, and substantial transition-metal $d$ character makes NbP a useful system for evaluating the interplay between surface topology and local chemisorption \rev{at} an electrocatalytic interface.

{\widowpenalty=10000
Here, we study hydrogen chemisorption on NbP(001) by combining density functional theory (DFT), Wannier-based tight-binding models, surface spectral functions, adsorbate--substrate bonding through the projected crystal orbital Hamilton population (pCOHP), and an adsorbate-projected current-induced spin-polarization analysis. SOC provides a controlled comparison between two electronic regimes of the same material: a nodal-line semimetal with drumhead-like surface states in the non-SOC limit and a Weyl semimetal with Fermi arcs in the presence of SOC. The lattice, surface termination, adsorption site, and local chemical structure are held fixed across this comparison.
\par}

Within this DFT comparison, the adsorption free energy and the hydrogen orbital occupation are nearly unchanged by SOC. The pCOHP analysis gives a net bonding contribution over the occupied energy window for the selected H-to-Nb/P subspaces, accumulated mainly below $E_F$; compensating Nb- and P-resolved changes leave the selected total adsorbate--substrate bonding nearly unchanged. The calculated SOC surface spectral functions retain the characteristic Fermi-arc branches after chemisorption and show H spectral weight in selected regions of the Fermi-level manifold. Together with the finite H-projected current-induced spin polarization, these results demonstrate the participation of the H-coupled spin-textured surface manifold in near-$E_F$ observables.



\section{Computational framework and electronic observables} \label{sec:ES}
\begin{revblock}
We compare matched non-SOC and SOC NbP(001) models, which support nodal-line-derived surface states and Weyl Fermi arcs, respectively.\cite{weng2015weyl,sun2015topological} Wannier projections resolve the bulk, surface, and adsorbate contributions used to analyze spectral weight, H--surface bonding, and current-induced spin polarization.
\end{revblock}

\subsection{Computational Details} \label{sec:Comp_D}
\begin{revblock}
We performed DFT simulations of NbP and NbP-H using VASP 5.4.4 \cite{kresse1996efficient,kresse1996efficiency} with the projector augmented-wave method \cite{blochl1994projector} and the PBE functional \cite{perdew1996generalized}. Spin-orbit coupling was treated self-consistently within the noncollinear formalism \cite{peralta2007noncollinear, kubler1988density, hobbs2000fully}. The geometries were relaxed with SOC and reused for the corresponding collinear non-SOC electronic calculations. A plane-wave energy cutoff of \(400\)~eV and unshifted \(8\times8\times1\) Monkhorst--Pack \(k\)-point meshes were employed \cite{monkhorst1976special}. The complete list of numerical parameters, including step-specific cutoffs and convergence criteria, is provided in Sec.~S1.1 of the Supplemental Information.
\end{revblock}

\noindent\textbf{Slab and adsorption geometry.}
NbP crystallizes in the noncentrosymmetric body-centered tetragonal structure with space group $I4_1md$ (No.~109) \cite{xu1996crystal}. The NbP(001) surface was modeled using an asymmetric 28 \rev{bilayer} slab with an Nb-terminated top surface and a P-terminated bottom surface, separated from periodic images by a vacuum region of 10~\AA. The slab preserves the stoichiometry of the bulk-truncated structure and exposes the Nb-rich surface relevant for hydrogen adsorption \cite{souma2016direct,sun2015topological}. During structural relaxation, the upper 20 \rev{bilayers} were allowed to relax while the remaining 8 lower \rev{bilayers} were kept fixed at their bulk-truncated positions.

\begin{center}
    \begin{figure}[t]
        \includegraphics[width=8.6cm]{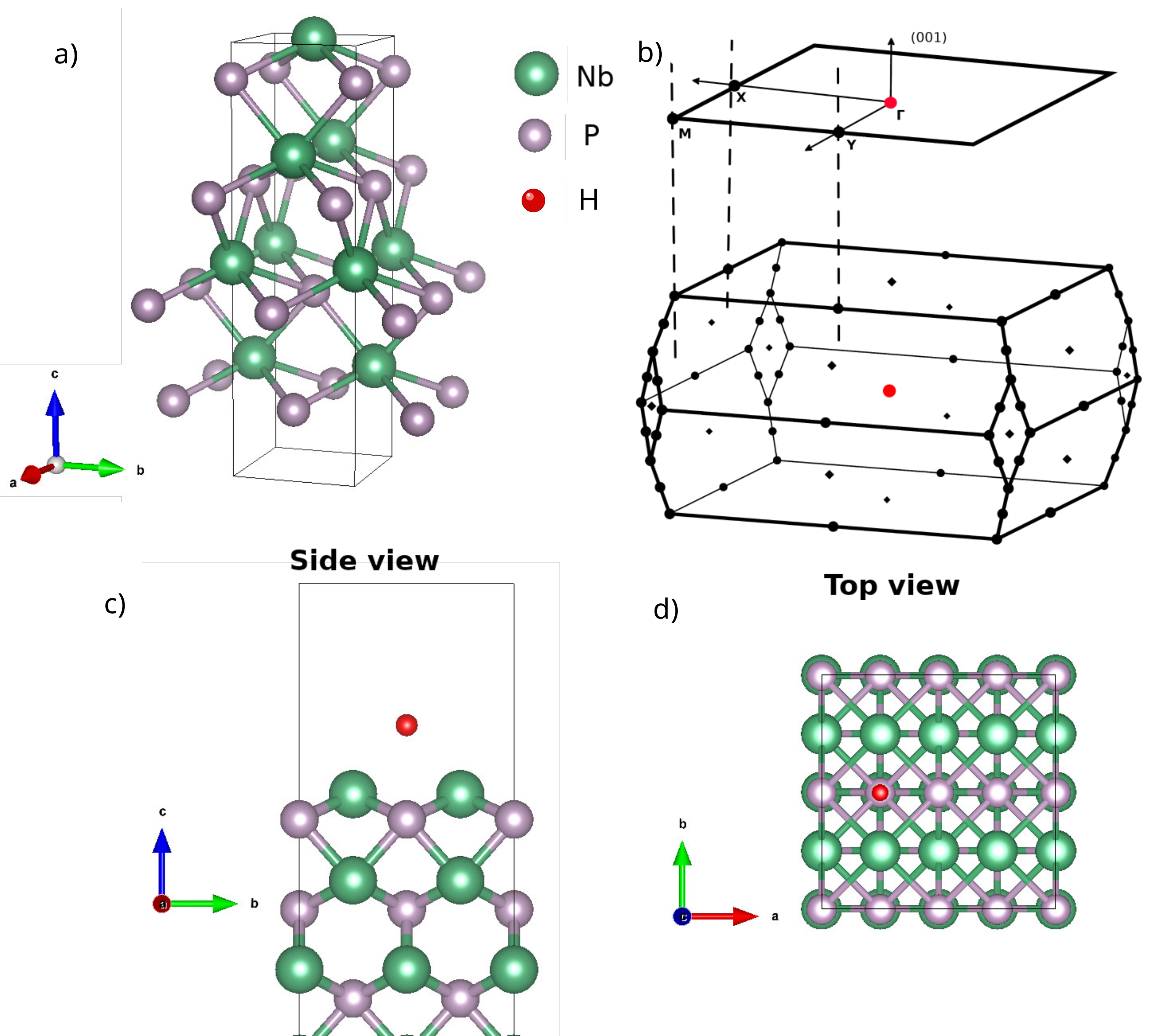}
        \caption{\rev{(a) Conventional NbP unit cell. (b) Primitive Brillouin zone projected along [001]. (c) Side view and (d) top view of the Nb-terminated NbP(001) surface with H adsorbed at the Nb--Nb bridge site. Green, lavender, and red spheres denote Nb, P, and H, respectively. The rectangle outlines the near-surface structural region; the surface projector used in the spectral analyses comprises the four outermost atomic layers (two Nb--P bilayers).}}
        \label{fig:prim_cell}
    \end{figure}
\end{center}

Hydrogen chemisorption was modeled by placing a single H layer above the Nb-terminated surface with one H atom included per primitive surface cell. \rev{Previous free-energy calculations for the transition-metal-monopnictide family identify bridge sites on Nb/Ta-terminated surfaces as the most active catalytic sites.\cite{rajamathi2017weyl} Because the noncentrosymmetric NbP(001) surface renders the two in-plane Nb--Nb bridge orientations inequivalent, we distinguish the \(x\)- and \(y\)-oriented bridge sites by their subsurface atomic registry. In the present structural relaxation, all three positional degrees of freedom of H were unconstrained, and the optimized structure places H above the \(y\)-oriented Nb--Nb bridge midpoint, as depicted in Fig.~\ref{fig:prim_cell}c,d. This optimized geometry was used for all subsequent H-covered calculations. Using the fixed in-plane lattice constant $a=3.35836$~\AA, the relaxed bridge-site geometry has an H adsorption height relative to the outermost Nb layer of $h_{\mathrm{H}}\approx 1.17$~\AA\ and two Nb--H bond lengths of $d_{\mathrm{Nb-H}}\approx 2.05$~\AA.} Additional structural parameters for this full-coverage model are reported in Table~S2 of the Supplemental Information.

\subsection{Hydrogen adsorption thermodynamics}
\label{sec:HER_thermo}

The Gibbs free energy of hydrogen adsorption, $\Delta G_{\mathrm{H}^\ast}$, is used as the thermodynamic descriptor for the equilibrium HER adsorption step. Within the computational hydrogen electrode approximation,\cite{skulason2010modeling}
\begin{equation}
    \Delta G_{\mathrm{H}^\ast} = \Delta E_{\mathrm{H}^\ast}^{\mathrm{ads}} + \Delta E_{\mathrm{ZPE}} - T\Delta S_{\mathrm{H}^\ast}.
    \label{eq:Gibbs}
\end{equation}
Here $\Delta E_{\mathrm{H}^\ast}^{\mathrm{ads}}$ is the DFT adsorption energy relative to the pristine slab and one half of gas-phase $\mathrm{H}_2$, while $\Delta E_{\mathrm{ZPE}}$ and $T\Delta S_{\mathrm{H}^\ast}$ are the zero-point-energy and entropy corrections. The detailed total-energy expression, vibrational correction, and entropy approximation are given in Sec.~S2 of the Supplemental Information.

\subsection{Wannier tight-binding Hamiltonian}
To analyze surface spectral weight and chemical bonding, we constructed tight-binding Hamiltonians from the electronic structure provided by the
DFT simulations using \textsc{Wannier90} \cite{mostofi2014updated}. The Wannier basis consists of MLWFs $\{|w_{\nu \mathbf{R}}\rangle\}$ obtained from projections onto Nb $d$ orbitals and P $s+p$ orbitals, with an additional H $1s$ orbital included for H-covered slabs \cite{marzari2012maximally}. The real-space Hamiltonian matrix elements are
\begin{equation}
    H_{\mu \nu}(\mathbf{R}) = \langle w_{\mu\mathbf{0}}| \hat{H}_{\mathrm{KS}} |w_{\nu\mathbf{R}}\rangle,
\end{equation}
where $\hat{H}_{\mathrm{KS}}$ is the Kohn--Sham Hamiltonian and $\mathbf{R}$ is a lattice vector connecting Wannier orbitals localized at
the corresponding unit cells. The Bloch Hamiltonian is obtained as
\begin{equation}
    H_{\mu \nu}(\mathbf{k}) = \sum_{\mathbf{R}} e^{i\mathbf{k}\cdot\mathbf{R}} H_{\mu \nu}(\mathbf{R}).
\end{equation}
This Wannier representation enables dense interpolation of the slab band structure and provides the localized orbital basis used for the projected spectral functions, pCOHP analysis, and electric field-induced spin-polarization response. Details of the Wannier projection and its validation against the DFT bands are reported in Secs.~S1.3 and S1.4 of the Supplemental Information.

\subsection{Projected spectral functions}
\label{sec:spectral_observables}

For a selected orbital or layer subspace \(\mathcal{P}\), we define the projected spectral function at energy \(E\) and in-plane crystal momentum \(\mathbf{k}_{\parallel}\) as
\begin{equation} \label{eq:pSF}
    A_P(\mathbf{k}_{\parallel},E) = \sum_n w^P_{n\mathbf{k}_{\parallel}} \, \delta_{\eta} \!\left(E-\varepsilon_{n\mathbf{k}_{\parallel}}\right),
\end{equation}
where the index \(n\) runs over the energy bands of the tight-binding Hamiltonian, $\delta_{\eta}$ is a broadened delta function, and
\begin{equation} \label{eq:Spectral_Weight}
    w^P_{n\mathbf{k}_{\parallel}} = \langle \psi_{n\mathbf{k}_{\parallel}}| \hat{P} |\psi_{n\mathbf{k}_{\parallel}}\rangle
\end{equation}
is the weight of eigenstate $|\psi_{n\mathbf{k}_{\parallel}}\rangle$ with energy \( \varepsilon_{n \mathbf{k}_{||}} \) in the projected subspace \( \mathcal{P} \). For surface spectral functions, $\hat{P}$ projects onto Wannier orbitals in the four outermost atomic layers (two Nb--P bilayers). For H-projected spectral functions, $\hat{P}$ projects onto the H $1s$ Wannier orbital.

\subsection{Crystal Orbital Hamilton Population}

We characterize selected adsorbate--surface orbital contributions using a Wannier-based projected Hamilton-population analysis related to COHP and pCOHP \cite{dronskowski1993crystal,deringer2011crystal,maintz2013analytic}. For each eigenstate, the implementation evaluates the phase-invariant H-to-target contribution $\operatorname{Re}[c_{\mu}^{*}H_{\mu\nu}c_{\nu}]$ once, with unit multiplicity. The reported quantities are finite-window bonding descriptors for the stated Wannier subspaces and orbital groups. Figures show $-\mathrm{pCOHP}$ and cumulative $-\overline{\mathrm{IpCOHP}}$, for which positive plotted values are assigned bonding character under this convention. The exact $\mathbf{k}_{\parallel}$ sampling, orbital blocks, validation tests, and integration are given in Sec. S4 of the Supplemental Information.

\subsection{Adsorbate-projected current-induced spin polarization}
\label{sec:edelstein_method}

The SOC calculation contains a spin-textured Fermi-level surface manifold, including branches associated with the established NbP(001) Fermi arcs \cite{sun2015topological}. Finite H $1s$ weight within this manifold provides an adsorbate-local channel for nonequilibrium spin accumulation under an in-plane electric field \( E_{\beta} \) along direction \(\beta\). This current-induced spin polarization is the projected interfacial form of the Edelstein response
\cite{edelstein1990spin, johansson2018edelstein}. We evaluate this response for the SOC calculations, where the spin-momentum-locked surface states provide the spin texture required for the Edelstein effect. The projected, area-normalized Pauli-spin response is written as
\begin{equation}
    \delta \sigma^{P,\mathrm{2D}}_{\alpha}=\sum_{\beta}\bar{\chi}^{P,\mathrm{2D}}_{\alpha\beta}E_{\beta},
\end{equation}
where $P$ denotes the selected local subspace, here the H $1s$ Wannier orbital or the surface Nb orbital block, and \(\bar{\chi}^{P,\mathrm{2D}}=\chi^P/(\lvert e\rvert A_{\mathrm{cell}})\) is reported in \(1/(\text{\AA}\,\mathrm{V})\). The corresponding projected current-induced magnetic-moment surface density is
\begin{equation}
    m^{P,\mathrm{2D}}_{\alpha}
    =
    -\mu_B
    \sum_{\beta}
    \bar{\chi}^{P,\mathrm{2D}}_{\alpha\beta}
    E_{\beta}.
    \label{eq:projected_moment}
\end{equation}
This quantity is the chemically resolved current-induced moment on the selected orbital subspace; projection over the full Wannier basis gives the corresponding total-slab quantity.

The detailed Kubo--Boltzmann expression, projected spin operator, and relaxation-time model are described in Sec.~S5 of the Supplemental Information.



\section{Results} \label{sec:Results}

\subsection{\rev{SOC dependence of adsorption thermodynamics}}
For H at the Nb--Nb bridge site, $\Delta E_{H^\ast}^{\mathrm{ads}}$ is $-0.83$~eV without SOC and $-0.82$~eV with SOC, while $\Delta G_{H^\ast}$ is $-0.62$ and $-0.61$~eV, respectively. The latter values are about 0.6~eV below thermoneutrality, indicating strong H overbinding. This full-coverage, solvent-free model uses one H per primitive cell at $0~\mathrm{V}_{\mathrm{RHE}}$ and pH~0; potential and coverage dependence and kinetics lie outside its scope.

The SOC-induced change in the adsorption free energy is below 0.02~eV, smaller than the exchange--correlation-functional variation and typical adsorption-energy errors reported in surface-adsorption benchmarks.\cite{wellendorff2012density,araujo2022adsorption} Accordingly, these calculations indicate that the adsorption thermodynamics is effectively insensitive to SOC within the accuracy of the present approach. Further discussion of functional sensitivity and the common slab electrostatic convention is provided in Supplemental Information Secs.~S1.1 and S2.

The VASP PAW/site-projected H \(1s\) occupation is approximately 0.51 electron in both cases, showing little SOC dependence within this local projector.

\begin{figure}[t]
    \includegraphics[width=\columnwidth]{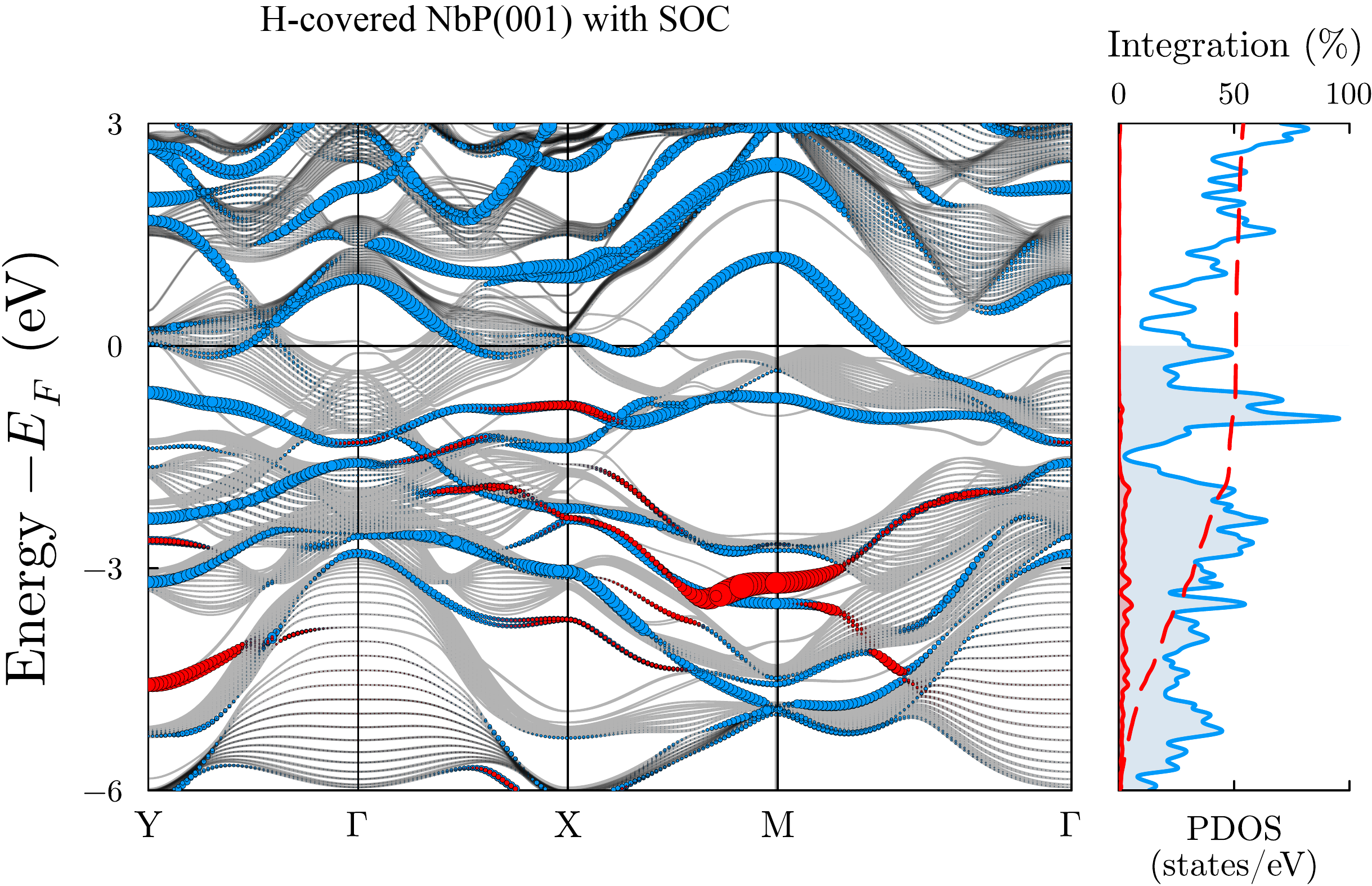}
    \caption{Representative SOC projected band structure for H-covered NbP(001). 
    Blue line thickness indicates localization on the four outermost atomic layers, 
    red color indicates H $1s$ spectral weight as defined in 
    Eq.~\eqref{eq:Spectral_Weight}, and faint gray curves show the remaining slab states. The right panel shows the VASP PAW/site-projected DOS: the solid blue curve is the combined surface-atom projection, the solid red curve is H$1s$, and the dashed red curve, read from the upper abscissa, is the cumulative H $1s$-projected DOS expressed as a percentage. Light shading marks the occupied contributions below $E_F$.
    The strongest H-derived spectral weight lies 
    mainly in occupied states below $E_F$;
    the corresponding non-SOC case is given in the Supplemental Information.}
    \label{fig:Bandstructure_H}
\end{figure}

Figure~\ref{fig:Bandstructure_H} shows the SOC H-projected 
band structure with color indicating H $1s$ weight 
and line thickness indicating surface localization.
The strongest H-derived spectral weight appears 
mainly below $E_F$, where occupied H--surface hybridized states form. 
Bands crossing at $E_F$ carry weaker and more selective H weight. 
The corresponding non-SOC projected band structure is 
similar and is given in the Supplemental Information. 
The projected bands place the occupied H-derived weight across a broad surface orbital manifold below $E_F$, including bulk-like and trivial contributions. We combine surface spectral functions that resolve the Fermi-level surface-state manifold with a projected Hamilton-population analysis of selected H-to-surface orbital contributions.

\subsection{\rev{H spectral weight within the Fermi-arc-bearing surface manifold at $E_F$}} \label{subsec:SF}

\begin{figure*}[t]
    \includegraphics[width=0.92\textwidth]{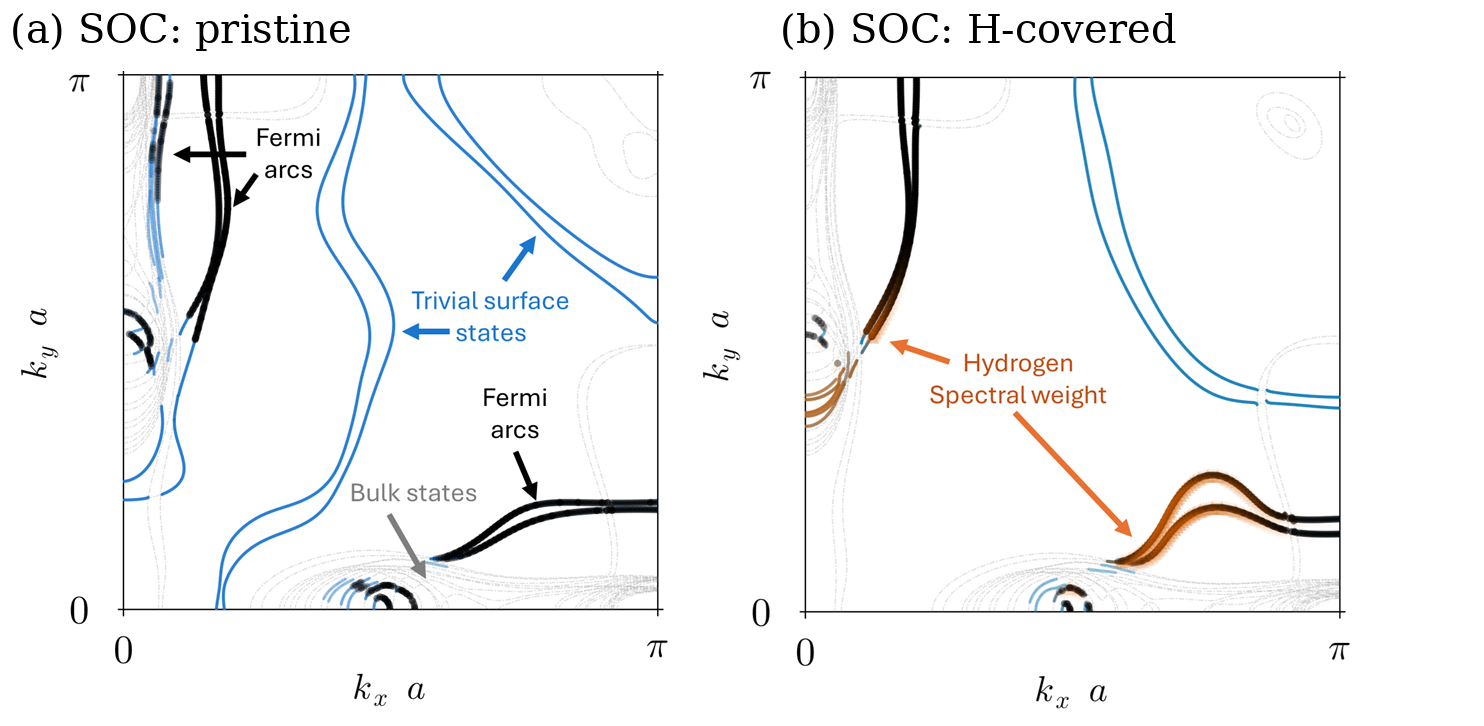}
    \caption{SOC surface spectral functions at $E_F$ for pristine and H-covered NbP(001). Black intensity marks surface spectral weight, blue contours mark additional surface-state features, gray contours indicate bulk-projected spectral weight, and orange intensity marks H-projected spectral weight in the H-covered calculation. The H-projected contribution uses a 20-fold display scale for graphical clarity. The characteristic NbP(001) Fermi-arc branches are retained in the H-covered calculation, and the H spectral weight is concentrated in selected momentum-space regions.}
    \label{fig:FA_highcov}
\end{figure*}

Figure~\ref{fig:FA_highcov} compares the SOC surface spectral functions at $E_F$ for pristine and H-covered NbP(001); the non-SOC maps are provided in the Supplemental Information. Guided by established NbP(001) surface-state assignments, the prominent Fermi-arc branches resolved in the pristine calculation are retained after H adsorption, and H-projected weight occurs in selected regions of the same Fermi-level manifold.

NbP hosts W$_1$ and W$_2$ Weyl nodes with termination-dependent surface states.\cite{souma2016direct,sun2015topological} Established NbP(001) analyses identify particularly clear Fermi-arc branches in momentum regions associated with the more widely separated W$_2$ projections.\cite{sun2015topological,belopolski2016criteria} The pristine/H-covered comparison therefore follows these prominent W$_2$-associated branches in the corresponding momentum-space regions.

\subsection{\rev{Momentum- and energy-resolved H--surface pCOHP}}
\label{sec:pCOHP_results}
\begin{figure*}[t]
    \includegraphics[width=0.53\textwidth]{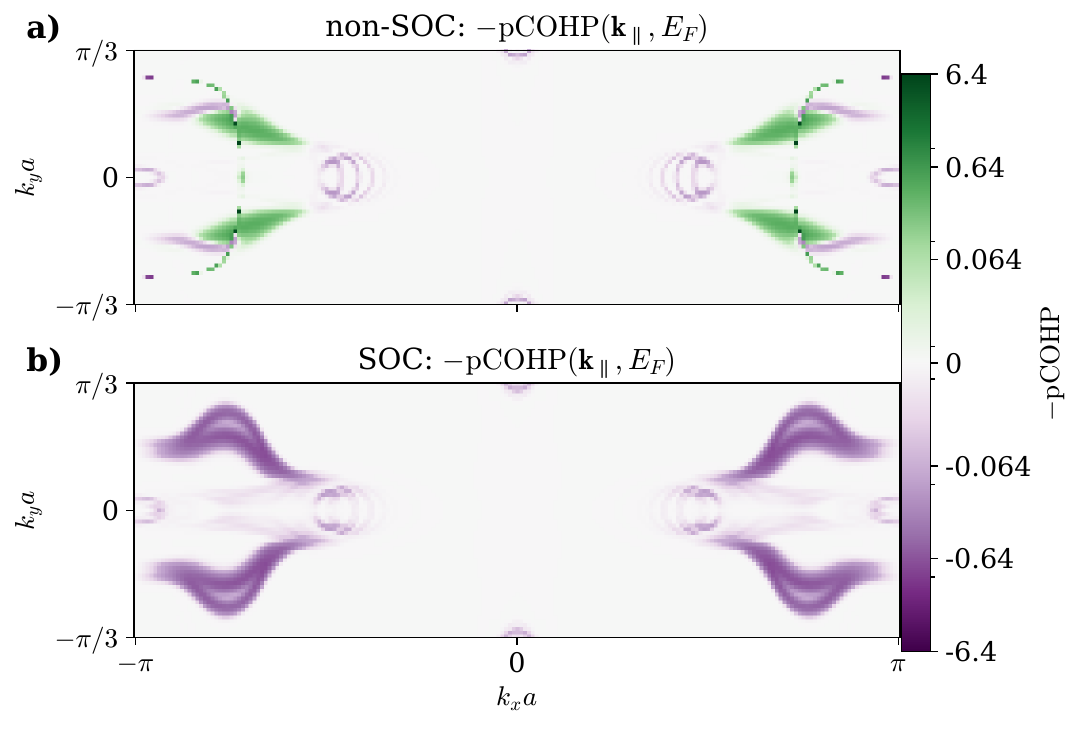}\hfill
    \includegraphics[width=0.43\textwidth]{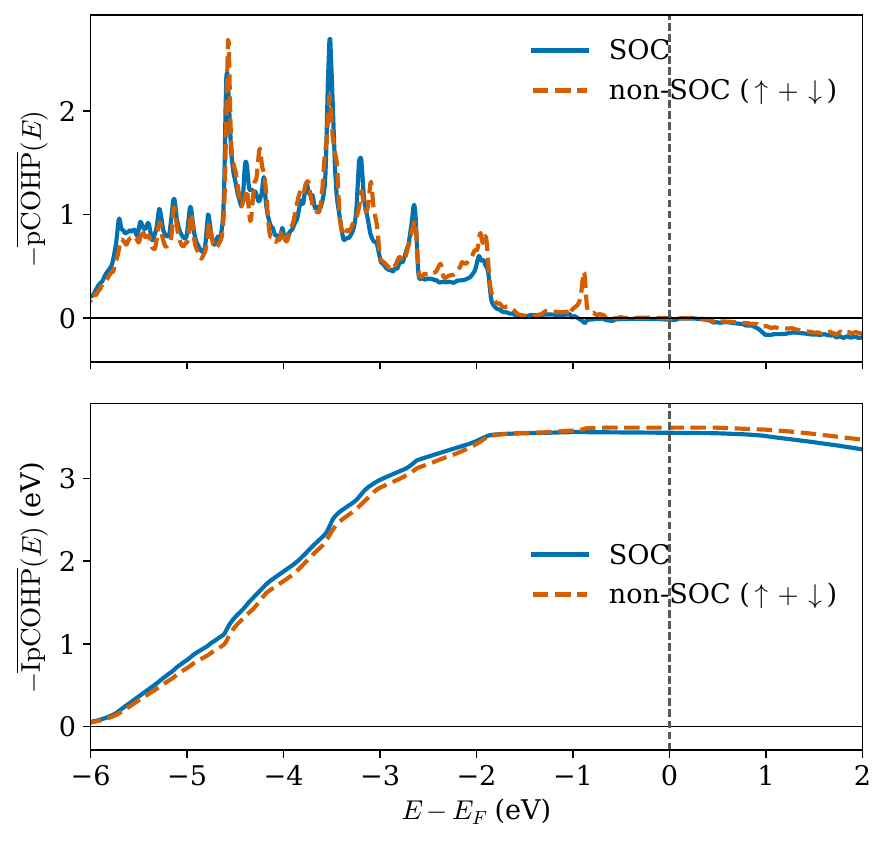}
    \caption{Momentum- and energy-resolved H--surface bonding populations for H-covered NbP(001). Left: $-\mathrm{pCOHP}(\mathbf{k}_{\parallel},E_F)$ between H $1s$ and the selected Nb1+Nb2+P1+P2 Wannier blocks without and with SOC, shown on a common color scale. Positive values denote bonding contributions and negative values antibonding contributions. The non-SOC calculation contains localized bonding features at $E_F$, whereas the largest SOC contributions are antibonding. Right: uniform-quadrant averages $-\overline{\mathrm{pCOHP}}(E)$ and cumulative $-\overline{\mathrm{IpCOHP}}(E)$. Both calculations yield a net bonding contribution over the occupied window, accumulated mainly below $E_F$. SOC redistributes the contributions among the selected Nb and P blocks while changing their total by only 1.8\%. Block-resolved results are provided in the Supplemental Information.}
    \label{fig:pCOHP_combined}
\end{figure*}

Figure~\ref{fig:pCOHP_combined} reports the projected Hamilton-population analysis between H $1s$ and selected Nb/P Wannier subspaces. The maps are evaluated directly at $E_F$. Under the plotted $-\mathrm{pCOHP}$ convention, the non-SOC map contains strong localized bonding features, whereas the largest SOC contributions are antibonding along the resolved momentum-space loci. The surface spectral functions in Fig.~\ref{fig:FA_highcov} display the prominent Fermi-arc branches and H-weighted momentum regions, complementing the momentum-resolved H-to-Nb/P bonding descriptors shown here. Together, these observables characterize the H-coupled surface manifold at $E_F$.

The energy-resolved curves give net positive occupied-window contributions for both electronic regimes, accumulated mainly well below $E_F$. On the common relative-energy grid, the selected Nb1+Nb2+P1+P2  value $-\overline{\mathrm{IpCOHP}}(E_F)$ changes from \(3.62\)~eV without SOC to \(3.55\)~eV with SOC. Here the labels Nb1, P1, Nb2, and P2 identify the atomic layers according to their position relative to the Nb-terminated surface. This 1.8\% change results from compensating block-resolved contributions: Nb1 increases from \(1.80\) to \(1.99\)~eV, whereas P1 decreases from \(1.74\) to \(1.49\)~eV; the Nb2 and P2 terms are much smaller. These finite-window H-to-target populations provide bonding descriptors for the selected subspaces.

\subsection{Current-induced spin polarization on chemisorbed H}
The SOC spectral functions show finite H $1s$  weight within the spin-textured Fermi-level surface manifold, including regions containing the resolved Fermi-arc features, motivating an adsorbate-projected current-induced spin-polarization analysis.
We evaluate the corresponding local magnetic-moment surface density
under in-plane electric fields using 
the projected response defined in Sec.~\ref{sec:edelstein_method},
\begin{equation}
    m_{\alpha}^{H,\mathrm{2D}}
    =
    -\mu_B
    \sum_{\beta}
    \bar{\chi}_{\alpha\beta}^{H,\mathrm{2D}}
    E_{\beta}.
\end{equation}
Here, \(\bar{\chi}^{H,\mathrm{2D}}=\chi^H/(\lvert e\rvert A_{\mathrm{cell}})\).

For an in-plane electric field of magnitude $10^5$~V/m, the strongest component of the H-projected current-induced moment reaches 
approximately $6\times10^{-7}\,\mu_B/\text{\AA}^2$. 
The H-covered surface-Nb projection is larger, 
reaching approximately $2.5\times10^{-6}\,\mu_B/\text{\AA}^2$, 
while the pristine surface-Nb reference reaches 
approximately $7.6\times10^{-6}\,\mu_B/\text{\AA}^2$. 
The smaller H-projected response is consistent with its lower spectral weight at $E_F$; the response remains finite and anisotropic.
Details of the projected response tensor computation
and the corresponding reference values
are described in Sec.~S5 in the Supplemental Information.



\section{Conclusions}
We examined how SOC-driven surface-state reorganization relates to H adsorption and adsorbate-local current-induced spin polarization on NbP(001). For the bridge-site geometry, the DFT adsorption free energies are nearly unchanged by SOC, while their values of \(-0.61\) to \(-0.62\)~eV indicate strong H overbinding relative to thermoneutrality.

The pCOHP analysis gives a net bonding contribution over the occupied window for the selected H-to-Nb/P subspaces, accumulated mainly below $E_F$. SOC strengthens the selected Nb1 contribution but weakens P1, changing the four-block total by only 1.8\%. The calculated SOC surface spectral functions retain the characteristic Fermi-arc branches after adsorption and show H $1s$ spectral weight in selected regions of the Fermi-level surface manifold. Together, the bonding and spectral analyses characterize the H-coupled surface-state manifold in near-$E_F$ H-projected observables.

The finite anisotropic H-projected current-induced spin polarization shows that chemisorbed hydrogen participates in the spin-textured surface response. NbP therefore provides a useful setting in which adsorption and electrically driven spin phenomena can be studied within the same topological surface manifold. Extending this approach across adsorbates, coverages, and electrochemical conditions may reveal how surface chemistry can be used to tune catalytic and spin-dependent interfacial responses together.


\hfill 

\par \textbf{Acknowledgments}
R.F.R. acknowledges generous start-up funds from Emory University. This work was supported by the U.S. Department of Energy, Office of Science, Office of Basic Energy Sciences, CPIMS program, under Award No. DE-SC0026024.\\

\par \textbf{Supplemental Information}
\rev{See the Supplemental Information for complete computational parameters and validation, adsorption and vibrational corrections, the slab electrostatic convention, and extended spectral, pCOHP, and current-induced-spin-response analyses.}

\bibliography{References}

@article{yan2017topological,
  title={Topological materials: {Weyl} semimetals},
  author={Yan, Binghai and Felser, Claudia},
  journal={Annual Review of Condensed Matter Physics},
  volume={8},
  number={1},
  pages={337--354},
  year={2017},
  publisher={Annual Reviews}
}

@article{weng2015weyl,
  title={{Weyl} semimetal phase in noncentrosymmetric transition-metal monophosphides},
  author={Weng, Hongming and Fang, Chen and Fang, Zhong and Bernevig, B Andrei and Dai, Xi},
  journal={Physical Review X},
  volume={5},
  number={1},
  pages={011029},
  year={2015},
  doi={10.1103/PhysRevX.5.011029},
  publisher={APS}
}

@article{xu2015discovery,
  title={Discovery of a {Weyl} fermion semimetal and topological {Fermi} arcs},
  author={Xu, Su-Yang and Belopolski, Ilya and Alidoust, Nasser and Neupane, Madhab and Bian, Guang and Zhang, Chenglong and Sankar, Raman and Chang, Guoqing and Yuan, Zhujun and Lee, Chi-Cheng and others},
  journal={Science},
  volume={349},
  number={6248},
  pages={613--617},
  year={2015},
  publisher={American Association for the Advancement of Science}
}

@article{lv2015experimental,
  title={Experimental discovery of {Weyl} semimetal {TaAs}},
  author={Lv, BQ and Weng, HM and Fu, BB and Wang, X Ps and Miao, Hu and Ma, Junzhang and Richard, P and Huang, XC and Zhao, LX and Chen, GF and others},
  journal={Physical Review X},
  volume={5},
  number={3},
  pages={031013},
  year={2015},
  publisher={APS}
}

@article{nielsen1983adler,
  title={The Adler-Bell-Jackiw anomaly and {Weyl} fermions in a crystal},
  author={Nielsen, Holger Bech and Ninomiya, Masao},
  journal={Physics Letters B},
  volume={130},
  number={6},
  pages={389--396},
  year={1983},
  publisher={Elsevier}
}

@article{kresse1996efficient,
  title={Efficient iterative schemes for ab initio total-energy calculations using a plane-wave basis set},
  author={Kresse, Georg and Furthm{\"u}ller, J{\"u}rgen},
  journal={Physical Review B},
  volume={54},
  number={16},
  pages={11169},
  year={1996},
  publisher={APS}
}

@article{kresse1996efficiency,
  title={Efficiency of ab-initio total energy calculations for metals and semiconductors using a plane-wave basis set},
  author={Kresse, Georg and Furthm{\"u}ller, J{\"u}rgen},
  journal={Computational Materials Science},
  volume={6},
  number={1},
  pages={15--50},
  year={1996},
  publisher={Elsevier}
}

@article{rajamathi2017weyl,
  title={{Weyl} semimetals as hydrogen evolution catalysts},
  author={Rajamathi, Catherine R. and Gupta, Uttam and Kumar, Nitesh and Yang, Hao and Sun, Yan and S{\"u}{\ss}, Vicky and Shekhar, Chandra and Schmidt, Marcus and Blumtritt, Horst and Werner, Peter and Yan, Binghai and Parkin, Stuart and Felser, Claudia and Rao, C. N. R.},
  journal={Advanced Materials},
  volume={29},
  number={19},
  pages={1606202},
  year={2017},
  doi={10.1002/adma.201606202},
  publisher={Wiley}
}

@article{murakami2007phase,
  title={Phase transition between the quantum spin Hall and insulator phases in {3D}: emergence of a topological gapless phase},
  author={Murakami, Shuichi},
  journal={New Journal of Physics},
  volume={9},
  number={9},
  pages={356},
  year={2007},
  publisher={IOP Publishing}
}

@article{asadchy2020sub,
  title={Sub-wavelength passive optical isolators using photonic structures based on {Weyl} semimetals},
  author={Asadchy, Viktar S and Guo, Cheng and Zhao, Bo and Fan, Shanhui},
  journal={Advanced Optical Materials},
  volume={8},
  number={16},
  pages={2000100},
  year={2020},
  publisher={Wiley Online Library}
}

@article{garcia2020optoelectronic,
  title={Optoelectronic response of the type-{I} {Weyl} semimetals {TaAs} and {NbAs} from first principles},
  author={Garcia, Christina AC and Coulter, Jennifer and Narang, Prineha},
  journal={Physical Review Research},
  volume={2},
  number={1},
  pages={013073},
  year={2020},
  publisher={APS}
}

@article{wang2018electron,
  title={Electron transport in {Dirac} and {Weyl} semimetals},
  author={Wang, Huichao and Wang, Jian},
  journal={Chinese Physics B},
  volume={27},
  number={10},
  pages={107402},
  year={2018},
  publisher={IOP Publishing}
}

@article{kharzeev2013anomaly,
  title={Anomaly induced chiral magnetic current in a {Weyl} semimetal: Chiral electronics},
  author={Kharzeev, Dmitri E and Yee, Ho-Ung},
  journal={Physical Review B},
  volume={88},
  number={11},
  pages={115119},
  year={2013},
  publisher={APS}
}

@article{shekhar2015extremely,
  title={Extremely large magnetoresistance and ultrahigh mobility in the topological {Weyl} semimetal candidate {NbP}},
  author={Shekhar, Chandra and Nayak, Ajaya K and Sun, Yan and Schmidt, Marcus and Nicklas, Michael and Leermakers, Inge and Zeitler, Uli and Skourski, Yurii and Wosnitza, Jochen and Liu, Zhongkai and others},
  journal={Nature Physics},
  volume={11},
  number={8},
  pages={645--649},
  year={2015},
  publisher={Nature Publishing Group UK London}
}

@article{hosur2012charge,
  title={Charge transport in {Weyl} semimetals},
  author={Hosur, Pavan and Parameswaran, SA and Vishwanath, Ashvin},
  journal={Physical Review Letters},
  volume={108},
  number={4},
  pages={046602},
  year={2012},
  publisher={APS}
}

@article{lundgren2014thermoelectric,
  title={Thermoelectric properties of {Weyl} and {Dirac} semimetals},
  author={Lundgren, Rex and Laurell, Pontus and Fiete, Gregory A},
  journal={Physical Review B},
  volume={90},
  number={16},
  pages={165115},
  year={2014},
  publisher={APS}
}

@article{sun2015topological,
  title={Topological surface states and {Fermi} arcs of the noncentrosymmetric {Weyl} semimetals {TaAs}, {TaP}, {NbAs}, and {NbP}},
  author={Sun, Yan and Wu, Shu-Chun and Yan, Binghai},
  journal={Physical Review B},
  volume={92},
  number={11},
  pages={115428},
  year={2015},
  doi={10.1103/PhysRevB.92.115428},
  publisher={APS}
}

@incollection{hammer2000theoretical,
  title={Theoretical surface science and catalysis—calculations and concepts},
  author={Hammer, Bj{\o}rk and N{\o}rskov, Jens Kehlet},
  booktitle={Advances in catalysis},
  volume={45},
  pages={71--129},
  year={2000},
  publisher={Elsevier}
}

@article{maintz2013analytic,
  title={Analytic projection from plane-wave and {PAW} wavefunctions and application to chemical-bonding analysis in solids},
  author={Maintz, Stefan and Deringer, Volker L and Tchougr{\'e}eff, Andrei L and Dronskowski, Richard},
  journal={Journal of computational chemistry},
  volume={34},
  number={29},
  pages={2557--2567},
  year={2013},
  publisher={Wiley Online Library}
}

@article{dronskowski1993crystal,
  title={Crystal orbital Hamilton populations ({COHP}): energy-resolved visualization of chemical bonding in solids based on density-functional calculations},
  author={Dronskowski, Richard and Bl{\"o}chl, Peter E.},
  journal={The Journal of Physical Chemistry},
  volume={97},
  number={33},
  pages={8617--8624},
  year={1993},
  publisher={ACS Publications},
  doi={10.1021/j100135a014},
}

@article{deringer2011crystal,
  title={Crystal orbital Hamilton population ({COHP}) analysis as projected from plane-wave basis sets},
  author={Deringer, Volker L and Tchougr{\'e}eff, Andrei L and Dronskowski, Richard},
  journal={The Journal of Physical Chemistry A},
  volume={115},
  number={21},
  pages={5461--5466},
  year={2011},
  publisher={ACS Publications},
  doi={10.1021/jp202489s},
}

@article{mostofi2014updated,
  title={An updated version of {Wannier90}: A tool for obtaining maximally-localised {Wannier} functions},
  author={Mostofi, Arash A and Yates, Jonathan R and Pizzi, Giovanni and Lee, Young-Su and Souza, Ivo and Vanderbilt, David and Marzari, Nicola},
  journal={Computer Physics Communications},
  volume={185},
  number={8},
  pages={2309--2310},
  year={2014},
  publisher={Elsevier}
}

@article{parsons1958rate,
  title={The rate of electrolytic hydrogen evolution and the heat of adsorption of hydrogen},
  author={Parsons, Roger},
  journal={Transactions of the Faraday Society},
  volume={54},
  pages={1053--1063},
  year={1958},
  publisher={Royal Society of Chemistry}
}

@article{trasatti1972work,
  title={Work function, electronegativity, and electrochemical behaviour of metals: {III}. Electrolytic hydrogen evolution in acid solutions},
  author={Trasatti, Sergio},
  journal={Journal of Electroanalytical Chemistry and Interfacial Electrochemistry},
  volume={39},
  number={1},
  pages={163--184},
  year={1972},
  publisher={Elsevier}
}

@article{wan2011topological,
  title={Topological semimetal and {Fermi}-arc surface states in the electronic structure of pyrochlore iridates},
  author={Wan, Xiangang and Turner, Ari M and Vishwanath, Ashvin and Savrasov, Sergey Y},
  journal={Physical Review B},
  volume={83},
  number={20},
  pages={205101},
  year={2011},
  publisher={APS}
}

@article{ruan2016symmetry,
  title={Symmetry-protected ideal {Weyl} semimetal in {HgTe}-class materials},
  author={Ruan, Jiawei and Jian, Shao-Kai and Yao, Hong and Zhang, Haijun and Zhang, Shou-Cheng and Xing, Dingyu},
  journal={Nature Communications},
  volume={7},
  number={1},
  pages={11136},
  year={2016},
  publisher={Nature Publishing Group UK London}
}

@article{weyl1929electron,
  title={{Elektron und Gravitation. I}},
  author={Weyl, Hermann},
  journal={Zeitschrift f{\"u}r Physik},
  volume={56},
  pages={330--352},
  year={1929},
  doi={10.1007/BF01339504}
}

@article{yang2015weyl,
  title={{Weyl} semimetal phase in the non-centrosymmetric compound {TaAs}},
  author={Yang, LX and Liu, ZK and Sun, Yan and Peng, Han and Yang, HF and Zhang, Teng and Zhou, Beatrice and Zhang, Yi and Guo, YF and Rahn, Marein and others},
  journal={Nature Physics},
  volume={11},
  number={9},
  pages={728--732},
  year={2015},
  publisher={Nature Publishing Group UK London}
}

@article{lv2015observation,
  title={Observation of {Weyl} nodes in {TaAs}},
  author={Lv, BQ and Xu, N and Weng, HM and Ma, JZ and Richard, P and Huang, XC and Zhao, LX and Chen, GF and Matt, CE and Bisti, F and others},
  journal={Nature Physics},
  volume={11},
  number={9},
  pages={724--727},
  year={2015},
  publisher={Nature Publishing Group UK London}
}

@article{jia2016weyl,
  title={{Weyl} semimetals, {Fermi} arcs and chiral anomalies},
  author={Jia, Shuang and Xu, Su-Yang and Hasan, M Zahid},
  journal={Nature Materials},
  volume={15},
  number={11},
  pages={1140--1144},
  year={2016},
  publisher={Nature Publishing Group UK London}
}

@article{xu2016spin,
  title={Spin polarization and texture of the {Fermi} arcs in the {Weyl} fermion semimetal {TaAs}},
  author={Xu, Su-Yang and Belopolski, Ilya and Sanchez, Daniel S and Neupane, Madhab and Chang, Guoqing and Yaji, Koichiro and Yuan, Zhujun and Zhang, Chenglong and Kuroda, Kenta and Bian, Guang and others},
  journal={Physical Review Letters},
  volume={116},
  number={9},
  pages={096801},
  year={2016},
  publisher={APS}
}

@article{souma2016direct,
  title={Direct observation of nonequivalent {Fermi}-arc states of opposite surfaces in the noncentrosymmetric {Weyl} semimetal {NbP}},
  author={Souma, S and Wang, Zhiwei and Kotaka, H and Sato, T and Nakayama, K and Tanaka, Y and Kimizuka, H and Takahashi, T and Yamauchi, K and Oguchi, T and others},
  journal={Physical Review B},
  volume={93},
  number={16},
  pages={161112},
  year={2016},
  publisher={APS}
}

@article{Sun2016,
  title={Strong Intrinsic {Spin Hall} Effect in the {TaAs} Family of {Weyl} Semimetals},
  author={Sun, Yan and Zhang, Yang and Felser, Claudia and Yan, Binghai},
  journal={Physical Review Letters},
  volume={117},
  number={14},
  pages={146403},
  year={2016},
  doi={10.1103/PhysRevLett.117.146403}
}

@article{perdew1996generalized,
  title={Generalized gradient approximation made simple},
  author={Perdew, John P and Burke, Kieron and Ernzerhof, Matthias},
  journal={Physical Review Letters},
  volume={77},
  number={18},
  pages={3865},
  year={1996},
  publisher={APS}
}

@article{araujo2022adsorption,
  title={Adsorption energies on transition metal surfaces: towards an accurate and balanced description},
  author={Araujo, Rafael B. and Rodrigues, Gabriel L. S. and dos Santos, Egon Campos and Pettersson, Lars G. M.},
  journal={Nature Communications},
  volume={13},
  number={1},
  pages={6853},
  year={2022},
  doi={10.1038/s41467-022-34507-y}
}

@article{wellendorff2012density,
  title={Density functionals for surface science: Exchange-correlation model development with Bayesian error estimation},
  author={Wellendorff, Jess and Lundgaard, Keld T. and M{\o}gelh{\o}j, Andreas and Petzold, Vivien G. and Landis, David D. and N{\o}rskov, Jens K. and Bligaard, Thomas and Jacobsen, Karsten W.},
  journal={Physical Review B},
  volume={85},
  number={23},
  pages={235149},
  year={2012},
  doi={10.1103/PhysRevB.85.235149}
}

@article{peralta2007noncollinear,
  title={Noncollinear magnetism in density functional calculations},
  author={Peralta, Juan E and Scuseria, Gustavo E and Frisch, Michael J},
  journal={Physical Review B},
  volume={75},
  number={12},
  pages={125119},
  year={2007},
  publisher={APS}
}

@article{kubler1988density,
  title={Density functional theory of non-collinear magnetism},
  author={Kubler, J and Hock, K-H and Sticht, J and Williams, AR},
  journal={Journal of Physics F: Metal Physics},
  volume={18},
  number={3},
  pages={469--483},
  year={1988}
}

@article{hobbs2000fully,
  title={Fully unconstrained noncollinear magnetism within the projector augmented-wave method},
  author={Hobbs, D and Kresse, G and Hafner, J},
  journal={Physical Review B},
  volume={62},
  number={17},
  pages={11556},
  year={2000},
  publisher={APS}
}

@article{monkhorst1976special,
  title={Special points for Brillouin-zone integrations},
  author={Monkhorst, Hendrik J and Pack, James D},
  journal={Physical Review B},
  volume={13},
  number={12},
  pages={5188},
  year={1976},
  publisher={APS}
}

@article{xu1996crystal,
  title={Crystal structure, electrical transport, and magnetic properties of niobium monophosphide},
  author={Xu, J and Greenblatt, M and Emge, T and H{\"o}hn, P and Hughbanks, T and Tian, Y},
  journal={Inorganic Chemistry},
  volume={35},
  number={4},
  pages={845--849},
  year={1996},
  publisher={ACS Publications}
}

@article{marzari2012maximally,
  title={Maximally localized {Wannier} functions: Theory and applications},
  author={Marzari, Nicola and Mostofi, Arash A and Yates, Jonathan R and Souza, Ivo and Vanderbilt, David},
  journal={Reviews of Modern Physics},
  volume={84},
  number={4},
  pages={1419--1475},
  year={2012},
  publisher={APS}
}

@article{johansson2024theory,
  title={Theory of spin and orbital {Edelstein} effects},
  author={Johansson, Annika},
  journal={Journal of Physics: Condensed Matter},
  volume={36},
  number={42},
  pages={423002},
  year={2024},
  publisher={IOP Publishing}
}

@article{aronov1989nuclear,
  title={Nuclear electric resonance and orientation of carrier spins by an electric field},
  author={Aronov, AG and Lyanda-Geller, Yu B},
  journal={Soviet Journal of Experimental and Theoretical Physics Letters},
  volume={50},
  pages={431},
  year={1989}
}

@article{dyakonov1971current,
  title={Current-induced spin orientation of electrons in semiconductors},
  author={Dyakonov, Mikhail I and Perel, VI},
  journal={Physics Letters A},
  volume={35},
  number={6},
  pages={459--460},
  year={1971},
  publisher={Elsevier}
}

@article{edelstein1990spin,
  title={Spin polarization of conduction electrons induced by electric current in two-dimensional asymmetric electron systems},
  author={Edelstein, Victor M},
  journal={Solid State Communications},
  volume={73},
  number={3},
  pages={233--235},
  year={1990},
  publisher={Elsevier}
}

@article{zeradjanin2016critical,
  title={A critical review on hydrogen evolution electrocatalysis: Re-exploring the volcano-relationship},
  author={Zeradjanin, Aleksandar R and Grote, Jan-Philipp and Polymeros, George and Mayrhofer, Karl JJ},
  journal={Electroanalysis},
  volume={28},
  number={10},
  pages={2256--2269},
  year={2016},
  publisher={Wiley Online Library}
}

@article{wu2020electronic,
  title={On the electronic structure and hydrogen evolution reaction activity of platinum group metal-based high-entropy-alloy nanoparticles},
  author={Wu, Dongshuang and Kusada, Kohei and Yamamoto, Tomokazu and Toriyama, Takaaki and Matsumura, Syo and Gueye, Ibrahima and Seo, Okkyun and Kim, Jaemyung and Hiroi, Satoshi and Sakata, Osami and others},
  journal={Chemical Science},
  volume={11},
  number={47},
  pages={12731--12736},
  year={2020},
  publisher={Royal Society of Chemistry}
}

@article{sarkar2018overview,
  title={An overview on {Pd}-based electrocatalysts for the hydrogen evolution reaction},
  author={Sarkar, Shreya and Peter, Sebastian C},
  journal={Inorganic Chemistry Frontiers},
  volume={5},
  number={9},
  pages={2060--2080},
  year={2018},
  publisher={Royal Society of Chemistry}
}

@article{hansen2021there,
  title={Is there anything better than {Pt} for {HER}?},
  author={Hansen, Johannes Novak and Prats, Hector and Toudahl, Karl Kr{\o}jer and M{\o}rch Secher, Niklas and Chan, Karen and Kibsgaard, Jakob and Chorkendorff, Ib},
  journal={ACS Energy Letters},
  volume={6},
  number={4},
  pages={1175--1180},
  year={2021},
  publisher={ACS Publications}
}

@article{johansson2018edelstein,
  title={{Edelstein} effect in {Weyl} semimetals},
  author={Johansson, Annika and Henk, J{\"u}rgen and Mertig, Ingrid},
  journal={Physical Review B},
  volume={97},
  number={8},
  pages={085417},
  year={2018},
  publisher={APS}
}

@article{johansson2016theoretical,
  title={Theoretical aspects of the {Edelstein} effect for anisotropic two-dimensional electron gas and topological insulators},
  author={Johansson, Annika and Henk, J{\"u}rgen and Mertig, Ingrid},
  journal={Physical Review B},
  volume={93},
  number={19},
  pages={195440},
  year={2016},
  publisher={APS}
}

@article{belopolski2016criteria,
  title={Criteria for directly detecting topological {Fermi} arcs in {Weyl} semimetals},
  author={Belopolski, Ilya and Xu, Su-Yang and Sanchez, Daniel S and Chang, Guoqing and Guo, Cheng and Neupane, Madhab and Zheng, Hao and Lee, Chi-Cheng and Huang, Shin-Ming and Bian, Guang and others},
  journal={Physical Review Letters},
  volume={116},
  number={6},
  pages={066802},
  year={2016},
  publisher={APS}
}

@article{blochl1994projector,
  title={Projector Augmented-Wave Method},
  author={Bl{\"o}chl, Peter E.},
  journal={Physical Review B},
  volume={50},
  number={24},
  pages={17953--17979},
  year={1994},
  doi={10.1103/PhysRevB.50.17953}
}

@article{yu2017nodal,
	title={From nodal chain semimetal to Weyl semimetal in HfC},
	author={Yu, Rui and Wu, Quansheng and Fang, Zhong and Weng, Hongming},
	journal={Physical review letters},
	volume={119},
	number={3},
	pages={036401},
	year={2017},
	publisher={APS}
}

@article{fang2016topological,
	title={Topological nodal line semimetals},
	author={Fang, Chen and Weng, Hongming and Dai, Xi and Fang, Zhong},
	journal={Chinese Physics B},
	volume={25},
	number={11},
	pages={117106},
	year={2016},
	publisher={IOP Publishing}
}

@article{medford2015sabatier,
	title={From the Sabatier principle to a predictive theory of transition-metal heterogeneous catalysis},
	author={Medford, Andrew J and Vojvodic, Aleksandra and Hummelsh{\o}j, Jens S and Voss, Johannes and Abild-Pedersen, Frank and Studt, Felix and Bligaard, Thomas and Nilsson, Anders and N{\o}rskov, Jens K},
	journal={Journal of Catalysis},
	volume={328},
	pages={36--42},
	year={2015},
	publisher={Elsevier}
}

@article{huang2015weyl,
	title={A Weyl Fermion semimetal with surface Fermi arcs in the transition metal monopnictide TaAs class},
	author={Huang, Shin-Ming and Xu, Su-Yang and Belopolski, Ilya and Lee, Chi-Cheng and Chang, Guoqing and Wang, BaoKai and Alidoust, Nasser and Bian, Guang and Neupane, Madhab and Zhang, Chenglong and others},
	journal={Nature communications},
	volume={6},
	number={1},
	pages={7373},
	year={2015},
	publisher={Nature Publishing Group UK London}
}

@article{jiao2022descriptors,
	title={Descriptors for the evaluation of electrocatalytic reactions: d-band theory and beyond},
	author={Jiao, Shilong and Fu, Xianwei and Huang, Hongwen},
	journal={Advanced Functional Materials},
	volume={32},
	number={4},
	pages={2107651},
	year={2022},
	publisher={Wiley Online Library}
}

@article{lee2015fermi,
	title={Fermi surface interconnectivity and topology in Weyl fermion semimetals TaAs, TaP, NbAs, and NbP},
	author={Lee, Chi-Cheng and Xu, Su-Yang and Huang, Shin-Ming and Sanchez, Daniel S and Belopolski, Ilya and Chang, Guoqing and Bian, Guang and Alidoust, Nasser and Zheng, Hao and Neupane, Madhab and others},
	journal={Physical Review B},
	volume={92},
	number={23},
	pages={235104},
	year={2015},
	publisher={APS}
}

@article{eftekhari2017electrocatalysts,
	title={Electrocatalysts for hydrogen evolution reaction},
	author={Eftekhari, Ali},
	journal={International Journal of Hydrogen Energy},
	volume={42},
	number={16},
	pages={11053--11077},
	year={2017},
	publisher={Elsevier}
}

@article{zheng2015advancing,
	title={Advancing the electrochemistry of the hydrogen-evolution reaction through combining experiment and theory},
	author={Zheng, Yao and Jiao, Yan and Jaroniec, Mietek and Qiao, Shi Zhang},
	journal={Angewandte Chemie International Edition},
	volume={54},
	number={1},
	pages={52--65},
	year={2015},
	publisher={Wiley Online Library}
}

@article{zhao2018heterostructures,
	title={Heterostructures for electrochemical hydrogen evolution reaction: a review},
	author={Zhao, Guoqiang and Rui, Kun and Dou, Shi Xue and Sun, Wenping},
	journal={Advanced Functional Materials},
	volume={28},
	number={43},
	pages={1803291},
	year={2018},
	publisher={Wiley Online Library}
}

@article{skulason2010modeling,
    title = {Modeling the electrochemical hydrogen oxidation and evolution reactions on the basis of density functional theory calculations},
    author = {Sk{\'u}lason, Egill and Tripkovic, Vladimir and Bj{\"o}rketun, M{\aa}rten E. and Gudmundsd{\'o}ttir, Sigr{\'\i}dur and Karlberg, Gustav and Rossmeisl, Jan and Bligaard, Thomas and J{\'o}nsson, Hannes and N{\o}rskov, Jens K.},
    journal = {The Journal of Physical Chemistry C},
    volume = {114},
    number = {42},
    pages = {18182--18197},
    year = {2010},
    publisher = {ACS Publications}
}

\end{document}